\documentclass[12pt,twoside]{article}
\usepackage{fleqn,espcrc1}

\usepackage{graphicx}
\usepackage{epsfig}
\usepackage[figuresright]{rotating}

\title{Light Hadron Spectroscopy at BEPC }

\author{Bing-Song Zou\address{CCAST (World Laboratory),
        P.O.~Box 8730, Beijing 100080
        and Institute of High Energy Physics, CAS,
        P.~O.~Box 918(4), Beijing 100039, P.R.China}}

\begin{document}

% typeset front matter
\maketitle

\begin{abstract}
The $J/\Psi$ and $\Psi'$ experiments at the Beijing Electron Positron 
Collider (BEPC) play a unique role in many aspects of light hadron
spectroscopy, such as hunting for glueballs and hybrids,
extracting $u\bar u+d\bar d$ and $s\bar s$ components of mesons,
and studying excited nucleons and hyperons, {\sl i.e.}, $N^*$, $\Lambda^*$, 
$\Sigma^*$ and $\Xi^*$ resonances. Physics objectives, recent results
and future prospects of light hadron spectroscopy at BEPC are presented.

\end{abstract}

%\begin{keyword}
%glueballs \sep hybrids \sep excite nucleons and hyperons
%\PACS 13.20.Gd \sep 14.40.Cs \sep 14.20.Gk \sep 14.20.Jn
%\end{keyword}

\section{Introduction}

There are two kinds of hadrons: mesons and baryons. They are the smallest
particles with sub-structure observed. Although we have already known for 
more than twenty years that hadrons are composed of quarks and gluons governed 
by strong interaction QCD, we still do not know how they are built up from 
these partons. The purpose of hadron spectroscopy is to explore the 
internal quark-gluon structure of the hadrons and the underlying strong
interaction QCD. It is a fundamental task for physicists. 

The Institute of High Energy Physics at Beijing runs an electron-positron
collider (BEPC) with a general purpose solenoidal detector, 
the BEijing Spectrometer (BES)\cite{BES}, 
which is designed to study exclusive final states in $e^+e^-$
annihilations at the center of mass energy from 2000 to 5600 MeV.
In this energy range, the largest cross sections are at the $J/\Psi(3097)$
and $\Psi'(3686)$ resonant peaks.
At present, the BES has collected about
30 million $J/\Psi$ events and 3.7 million $\Psi'$ events. More data are
going to be taken. From $J/\Psi$ and $\Psi'$ decays, both meson spectroscopy
and baryon spectroscopy can be studied. 

Three main processes which play
a unique role for the light hadron spectroscopy are $\Psi$ radiative decay,
$\Psi$ hadronic decay into mesons, and $\Psi$ hadronic decay into baryons 
and anti-baryons. In the following three sections, I will outline the
physics objectives and summarize recent results for each of them. 
The outlook is given in the final section.

\section{$\Psi$ radiative decays}

\begin{figure}[htbp]
\vspace{-1.6cm}
\hspace{-1.5cm}\includegraphics[width=17cm,height=7cm]{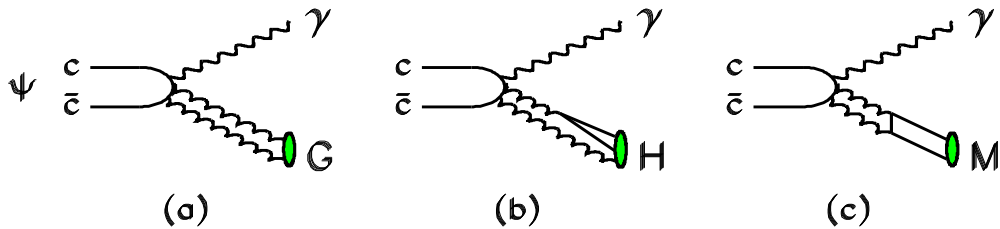}   
\vspace{-1.8cm}
\caption{$\Psi$ radiative decays to (a) glueball, (b)
hybrid, and (c) $q\bar q$ meson. }
\label{fig:1}
\end{figure}

There are three main physics objectives for {$\Psi$ radiative decays:

(1) Looking for glueballs and hybrids. As shown in Fig.~\ref{fig:1},
after emitting a photon, the $c\bar c$ pair is in a $C=+1$ state and
decays to hadrons dominantly through two gluon intermediate states.
Simply counting the power of $\alpha_s$ we know that glueballs should
have the largest production rate, hybrids the second,
then the ordinary $q\bar q$ mesons.

(2) Completing $q\bar q$ meson spectroscopy and studying their production
and decay rates, which is crucial for understanding their internal
structure and confinement.

(3) Extracting $gg\leftrightarrow q\bar q$ coupling from perturbative
energy region of above 3 GeV to nonperturbative region of 0.3 GeV.
This may show us some phenomenological pattern for the smooth transition
from perturbative QCD to strong nonperturbative QCD.

Up to now, we have mainly worked on glueball searches. One thing worth
noting is that the $J/\Psi$ radiative decay has a similar decay pattern as 
$0^{-+}$, $0^{++}$ and $2^{++}$ charmoniums, {\sl i.e.}, $\eta_c$,
$\chi_{c0}$ and $\chi_{c2}$, as it should be, since all of them decay
through two gluons. The $4\pi$, $\bar KK\pi\pi$, $\eta\pi\pi$ and 
$\bar KK\pi$ seem to be the most favorable final states for the two gluon
transition. The branching ratios for $J/\Psi$ radiative decay to these
four channels are listed in Table \ref{tab1}. The sum of them is about
half of all radiative decays.  If glueballs exist, they should appear in
these four channels. Therefore BES Collaboration has recently performed
partial wave analyses (PWA) of these four
channels\cite{BES1,BES2,BES3,BES4} as
well as $\gamma\bar KK$ channel\cite{BES5,Shen}. The hadronic invariant
mass spectra for these channels are shown in
Figs.~\ref{fig2k2p}-\ref{figkk}.

\begin{table}[htb]
\caption{Branching ratios for the four largest $J/\Psi$ radiative decay
channels (BR$\times 10^3$)}
\label{tab1}
\renewcommand{\arraystretch}{1.2} %enlarge line spacing
\begin{center}
\begin{tabular}{cccc} 
\hline
$\gamma 4\pi$ & $\gamma\bar KK\pi\pi$ & $\gamma\eta\pi\pi$ &
$\gamma\bar KK\pi$\\
\hline
$14.4\pm 1.8$ \cite{Kopke} & $9.5\pm 2.7$ \cite{BES2} 
& $6.1\pm 1.0$ \cite{PDG} & $6.0\pm 2.1$ \cite{BES4} \\
\hline
\end{tabular}\\
\end{center}
\end{table}

\begin{figure}[htbp]
\vspace{-1.5cm}
\hspace{-0.5cm}\includegraphics[width=17cm,height=7cm]{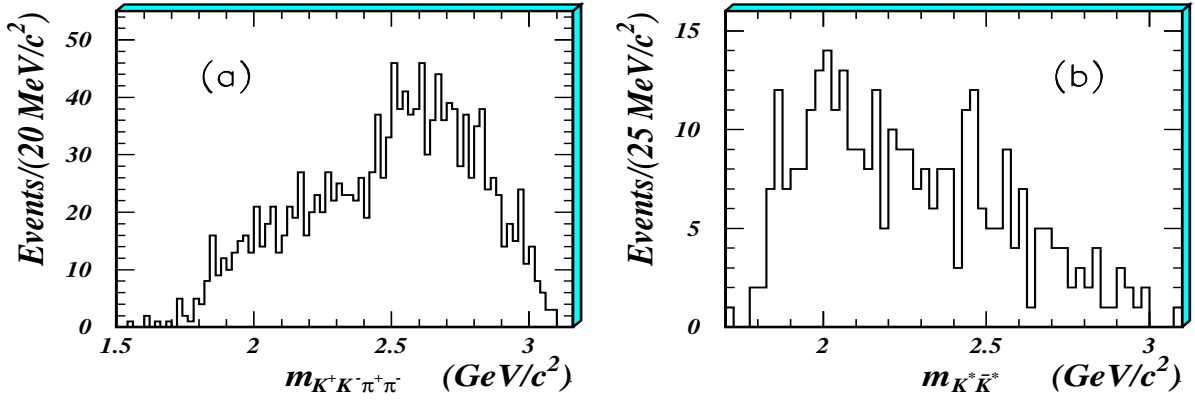}   
\vspace{-1.5cm}
\caption{(a) The $KK\pi\pi$ mass of $J/\Psi\to\gamma K^+K^-\pi^+\pi^-$; 
(b) The $K^*\bar K^*$ mass of $J/\Psi\to\gamma K^*\bar K^*$.\cite{BES2} }
\label{fig2k2p}
\end{figure}

\begin{figure}[htbp] 
\vspace{-0.3cm}
\begin{minipage}[t]{70mm}
\centerline{\epsfig{file=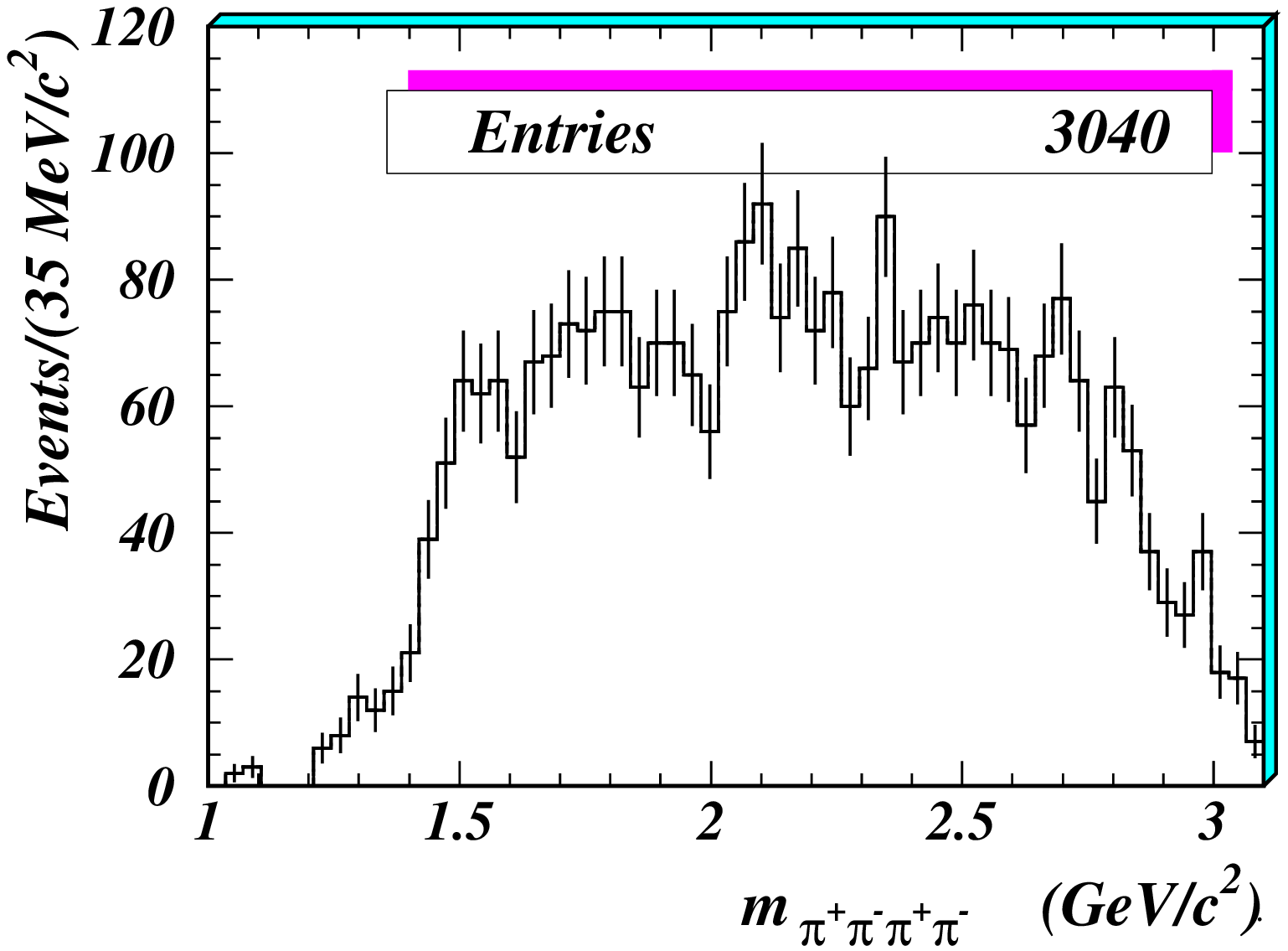,height=6.5cm,width=7cm}}
\vspace{-1.cm}\caption[]{The $4\pi$ mass of
$J/\Psi\to\gamma\pi^+\pi^-\pi^+\pi^-$.\cite{BES1}}
\label{fig4pi}
\end{minipage}
\hspace{\fill}
\begin{minipage}[t]{70mm}
\centerline{\epsfig{file=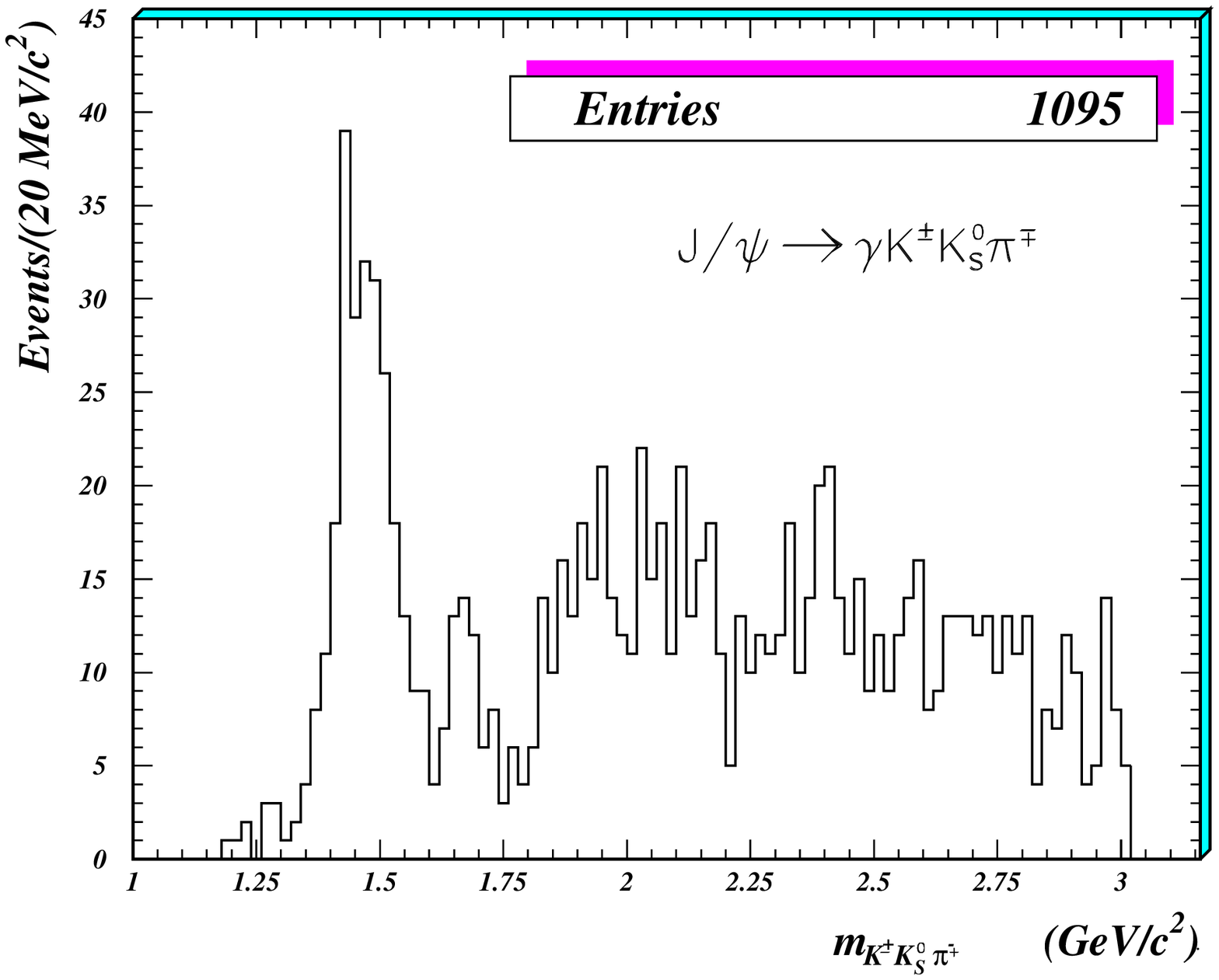,height=6.5cm,width=7cm}}
\vspace{-1.cm}\caption[]{The $K_SK^\pm\pi^\mp$ mass of $J/\Psi\to\gamma 
K_SK^\pm\pi^\mp$.\cite{BES4}}
\label{figkkp}
\end{minipage}  
\end{figure}

\begin{figure}[htbp] 
\vspace{-0.3cm}
\begin{minipage}[t]{70mm}
\hspace{4mm}\centerline{\epsfig{file=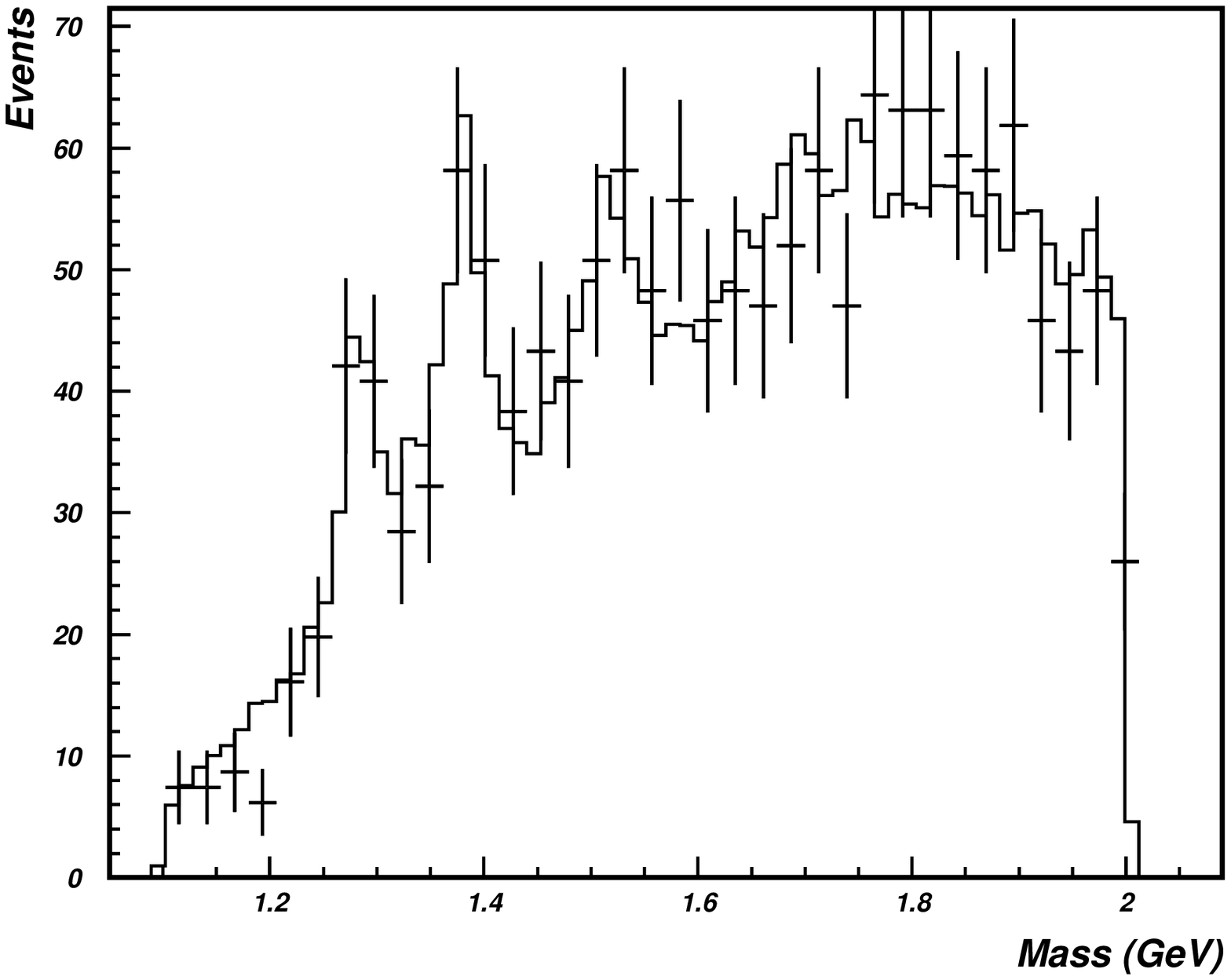,height=6.5cm,width=7cm}}
\vspace{-1.3cm}\caption[]{The $\eta\pi\pi$ mass of
$J/\Psi\to\gamma\eta\pi^+\pi^-$.\cite{BES3}}
\label{figpep}
\end{minipage}
\hspace{\fill}
\begin{minipage}[t]{70mm}
\centerline{\epsfig{file=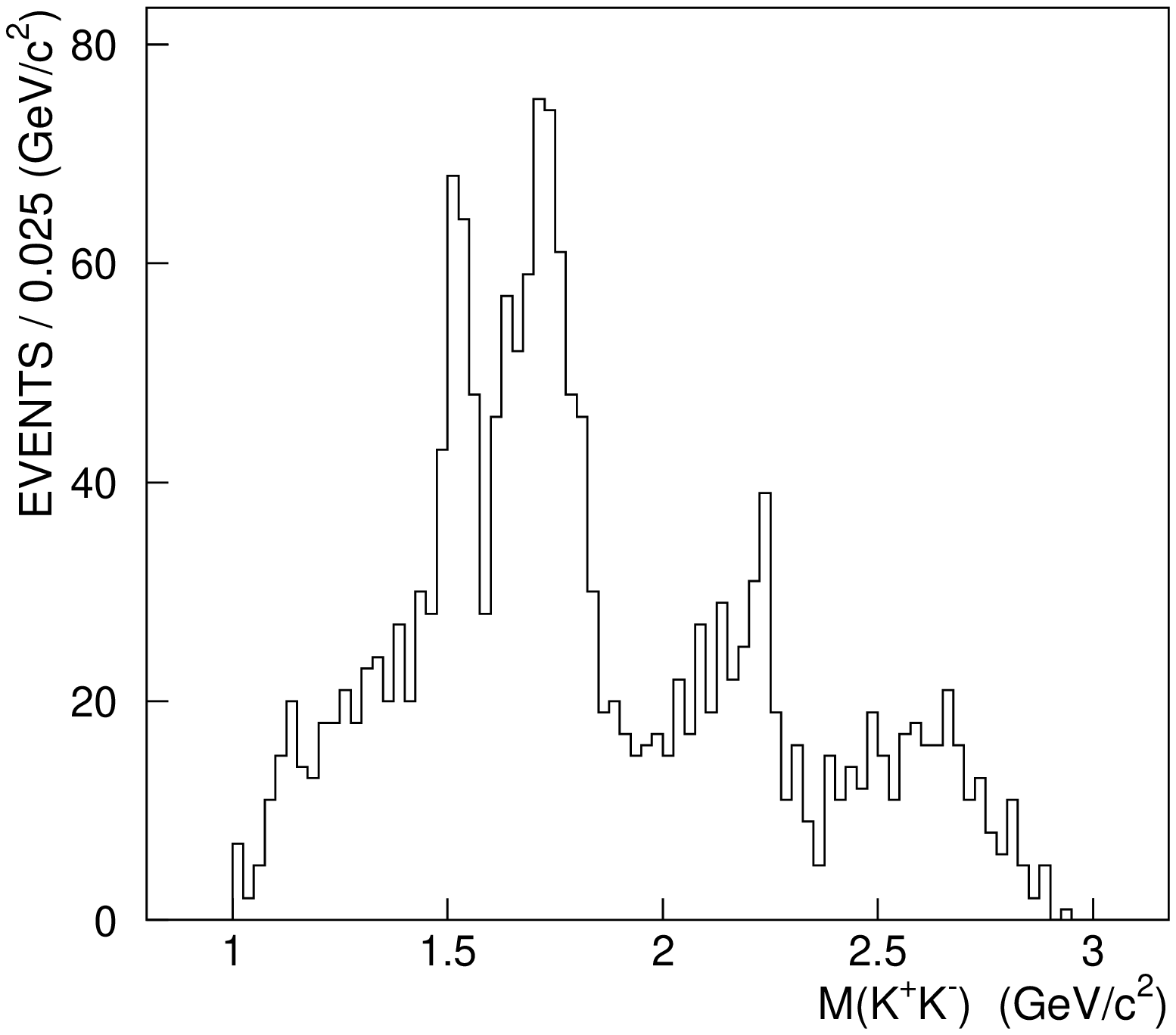,height=5.7cm,width=7cm}}
\vspace{-0.8cm}
\caption[]{The $K\bar K$ mass of $J/\Psi\to\gamma K^+K^-$.\cite{BES5}}
\label{figkk}
\end{minipage}  
\end{figure}

The BES results on $J/\Psi\to\gamma\pi^+\pi^-\pi^+\pi^-$\cite{BES1} are
approximately consistent with the results from the earlier re-analysis of
MARKIII data\cite{Scott} where three $0^{++}$ resonances are needed at
(1500, 1750, 2100) MeV decaying dominantly through $\sigma\sigma$
intermediate state, plus a very broad $0^{-+}$ component named
$\eta(1800)$ first\cite{BES3} and $\eta(2190)$ later\cite{BDZ}. But the
BES data favor more $2^{++}$ in the high
mass region. A broad $2^{++}$ resonance $f_2(1950)$ is needed around 2 GeV
together with some contribution from $f_2(1270)$ and $f_2(1565)$.

For $J/\Psi\to\gamma K^+K^-\pi^+\pi^-$\cite{BES2}, we found that the
$K^*\bar K^*$ contribution peaks strongly near threshold, which can be
fitted with a broad $0^{-+}$ resonance compatible with
$\eta(2190)$\cite{BDZ}. The data also definitely required a broad $2^{++}$
resonance $f_2(1950)$ decaying to $K^*\bar K^*$. There is further evidence
for a $2^{-+}$ component peaking at 2.55 GeV. The non-$K^*\bar K^*$
contribution is close to phase space; it peaks at 2.6 GeV and is very
different from $K^*\bar K^*$.

For $J/\Psi\to\gamma\eta\pi^+\pi^-$\cite{BES3}, the main discovery from
the BES data is that the $0^{-+}$ contribution dominates with three
$0^{-+}$ resonances, $\eta(1440)$, $\eta(1760)$ and $\eta(2190)$ (named
$\eta(1800)$ at that time). In addition, there is a definite $2^{-+}$ 
signal for the $\eta_2(1870)$ decaying dominantly to $f_2\sigma$.

The structures in $J/\Psi\to\gamma K^+K^-\pi^0$ and $K_SK^\pm\pi^\mp$ 
\cite{BES4} are relatively simple. Two largest components are 
from $0^{-+}$ resonances $\eta(1440)$ and $\eta(2040)$.

We also re-analyzed the $J/\Psi\to\gamma K^+K^-$ data, $0^{++}$
contribution ($75\%\sim 80\%$) dominates the $f_J(1710)$
peak\cite{BES5,Shen}.
Previous moment analysis suffers a problem of incomplete detection phase
space\cite{Chen1}.

From these analyses together with information from
other sources\cite{PDG,Chen2}, we get $J/\Psi$ radiative decay branching
ratios for the production of $0^{-+}$, $2^{++}$, $0^{++}$ and $2^{-+}$
resonances of $1\sim 2.5$ GeV as listed in Table\ref{tab2}.

\begin{table}[htb]
\caption{$J/\Psi$ radiative decay branching ratios for the production of
$0^{-+}$, $2^{++}$, $0^{++}$ and $2^{-+}$ resonances of $1\sim 2.5$ GeV
(BR$\times 10^3$)}
\label{tab2}
\renewcommand{\arraystretch}{1.2} %enlarge line spacing
\begin{center}
\begin{tabular}{cccccc} 
\hline
$\eta(2190)$ & $19\pm 3$ \cite{BPZ,BES1,BES2,BES3,BES4} & $f_2(1950)$ &
$2.6\pm 0.6$\cite{BPZ,BES1,BES2} & $f_0(1500)$ & $1.1\pm 0.1$
\cite{BES1,PDG}\\
$\eta(1440)$ & $2.1\pm 0.7$ \cite{BES3,BES4} & $f_2(1270)$ &
$1.4\pm 0.1$\cite{PDG} & $f_0(1740)$ & $1.1\pm 0.1$
\cite{BES1,BES5}\\
$\eta(2040)$ & $2.1\pm 0.7$ \cite{BES4} & $f_2(1560)$ &
$1.0\pm 0.1$\cite{BES1} & $f_0(2100)$ & $0.7\pm 0.2$
\cite{BES1}\\
$\eta(1760)$ & $1.8\pm 0.8$ \cite{BES3} & $f'_2(1525)$ &
$0.5\pm 0.1$\cite{PDG} & $\eta_2(1870)$ & $0.9\pm 0.3$
\cite{BES3}\\
\hline
\end{tabular}\\
\end{center}
\end{table}

The largest branching ratio of the $J/\Psi$ radiative decay is for the
very broad $0^{-+}$ resonance $\eta(2190)$ with a full width of 
$(850\pm 100)$ MeV. It decays to $\rho\rho$, $\omega\omega$, 
$K^*\bar K^*$ and $\phi\phi$ in a flavor-blind way within
errors\cite{BDZ}. These properties suggest that it is the $0^-$ glueball
predicted near this mass by lattice calculations. Its mixing with
other $0^{-+}$ resonances will give a natural explanation for the large
production rates for all the $0^{-+}$ resonances listed in the
Table\ref{tab2}, which has long been known as the pseudoscalar
puzzle\cite{Kopke}.

The largest $2^{++}$ production rate is for the broad $f_2(1950)$
decaying to $4\pi$\cite{BES1} and $K^*\bar K^*$\cite{BES2}.
It also appears in two other glue-rich processes: proton-proton central
production\cite{Kirk} and proton-antiproton annihilation\cite{CB1950}.
Especially, it has a much larger production rate\cite{Kirk} in the
central production process than other nearby $2^{++}$ resonances above 1.5
GeV. These properties make the $f_2(1950)$ the best $2^{++}$ glueball
candidate\cite{BPZ}. 

The two $0^{++}$ resonances, $f_0(1500)$ and $f_0(1740\pm 30)$, have
similar production rates in the $J/\Psi$ radiative decays. However, the
$f_0(1740\pm 30)$ has much weaker production rates than the $f_0(1500)$ in
two other glue-rich processes\cite{Kirk,CB1950} and much stronger
production rates in glueball disfavoring processes such as the
photon-photon collision\cite{Braccini} and the $J/\Psi\to\omega K^+K^-$ 
process\cite{DM2}. This makes the $f_0(1500)$ the best $0^{++}$ glueball
candidate. But it is possible that the bare glueball state is dispersed
over three real resonances, $f_0(1500)$, $f_0(1740)$ and
the nearby broad state $f_0(1670)$\cite{LZL} or $f_0(1530)$\cite{ASA} or
$f_0(1370)$\cite{AC}. 

In summary, from our partial wave analyses of the four largest $J/\Psi$
radiative decays together with information from other sources, we find
that $f_0(1500)$, $\eta(2190)$ and $f_2(1950)$ display the strongest
glueball characteristics. Their mass ratios agree well with the latest
predictions of Lattice Gauge calculations of the glueball
spectrum\cite{BPZ}.

As for the isoscalar $1^{-+}$ hybrids, they are predicted to decay
dominantly into $4\pi$ through $a_1\pi$ and $\pi(1300)\pi$ intermediate
states\cite{PageH}. If they exist, they should appear in the
$J/\Psi\to\gamma\pi^+\pi^-\pi^+\pi^-$ process. In our present partial
wave analysis of this channel, we have not include this exotic quantum
number. We shall re-analyze this channel by allowing this possibility
after more statistics on this channel is available. 

A special subsector of $J/\Psi$ radiative decays, 
$J/\Psi\to\gamma A\to\gamma(\gamma\rho : \gamma\omega : \gamma\phi)$, 
is on our list for the partial wave analysis in near future.
This process filters the flavor content of all $C = +$ states and
disfavors the glueball production since glueballs do not couple to
the photon directly. BES data for $J/\Psi\to\gamma(\gamma\rho :
\gamma\phi)$ show clear evidence for the $\eta(1440)$ decaying to
$\gamma\rho$ and $\gamma\phi$ \cite{XUGF}. 

\section{$\Psi$ hadronic decays to mesons}

\begin{figure}[htbp]
\vspace{-1.6cm}
\hspace{-1.5cm}\includegraphics[width=17cm,height=7cm]{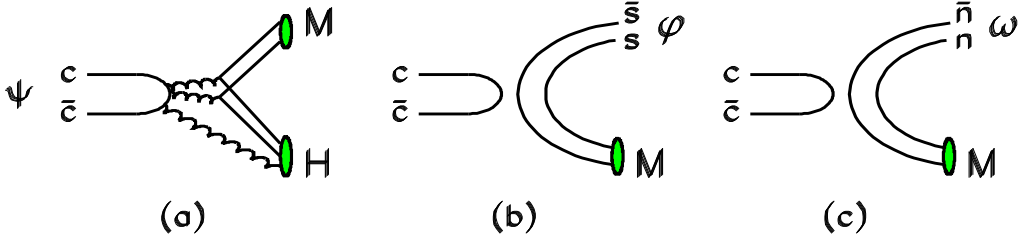}   
\vspace{-1.8cm}
\caption{$\Psi$ hadronic decays to (a) hybrids, (b)
$s\bar s$, and (c) $n\bar n\equiv {1\over\sqrt{2}}(u\bar u+d\bar d)$
mesons. }
\label{fig:2}
\end{figure}

There are also mainly three physics objectives here:

(1) Looking for hybrids. Since $\Psi$ decays to hadrons through three
gluons, final states involving a hybrid as shown in Fig.~\ref{fig:2}(a) 
are expected to have larger production rate than ordinary $q\bar q$ mesons
as shown in Fig.~\ref{fig:2}(b,c).

(2) Extracting $u\bar u+d\bar d$ and $s\bar s$ components of
associated mesons, M, via $\Psi\to M+\omega/\phi$ as shown in
Fig.~\ref{fig:2}(b,c).

(3) Disfavoring glueball production. We can analyze the quark/gluon
content of a particle, M, by comparing its production in 
$\Psi\to\gamma M$, $\omega M$ and $\phi M$.

In order to look for isoscalar $1^{-+}$ hybrid $\hat\omega$ decaying to
$4\pi$, we have studied $J/\Psi\to\omega\pi^+\pi^-\pi^+\pi^-$
process\cite{Zhu}. A peak around 1.75 GeV in the $4\pi$ invariant mass
spectrum is visible. But due to low statistics, no PWA is performed.
No other structure is observed. 

To investigate the $u\bar u+d\bar d$ and $s\bar s$ components of mesons,
we have studied $J/\psi\to\omega\pi^+\pi^-$, $\omega K^+K^-$,
$\phi\pi^+\pi^-$ and $\phi K^+K^-$ channels. The invariant mass spectra
for these channels are similar to the previous ones by MARKIII and DM2
Collaborations.

For $J/\psi\to\omega\pi^+\pi^-$, there are two clear peaks at 500 MeV and
1275 MeV in the $2\pi$ mass spectrum corresponding to the $\sigma$ 
and the $f_2(1275)$, respectively\cite{Shen}. 
For $J/\psi\to\omega K^+K^-$, there
is a threshold enhancement due to the $f_0(980)$ and a clear peak at 1710
MeV probably due to the $f_0(1710)$.

\begin{figure}[htbp] 
\vspace{-1.cm}
\begin{minipage}[t]{75mm}
\centerline{\epsfig{file=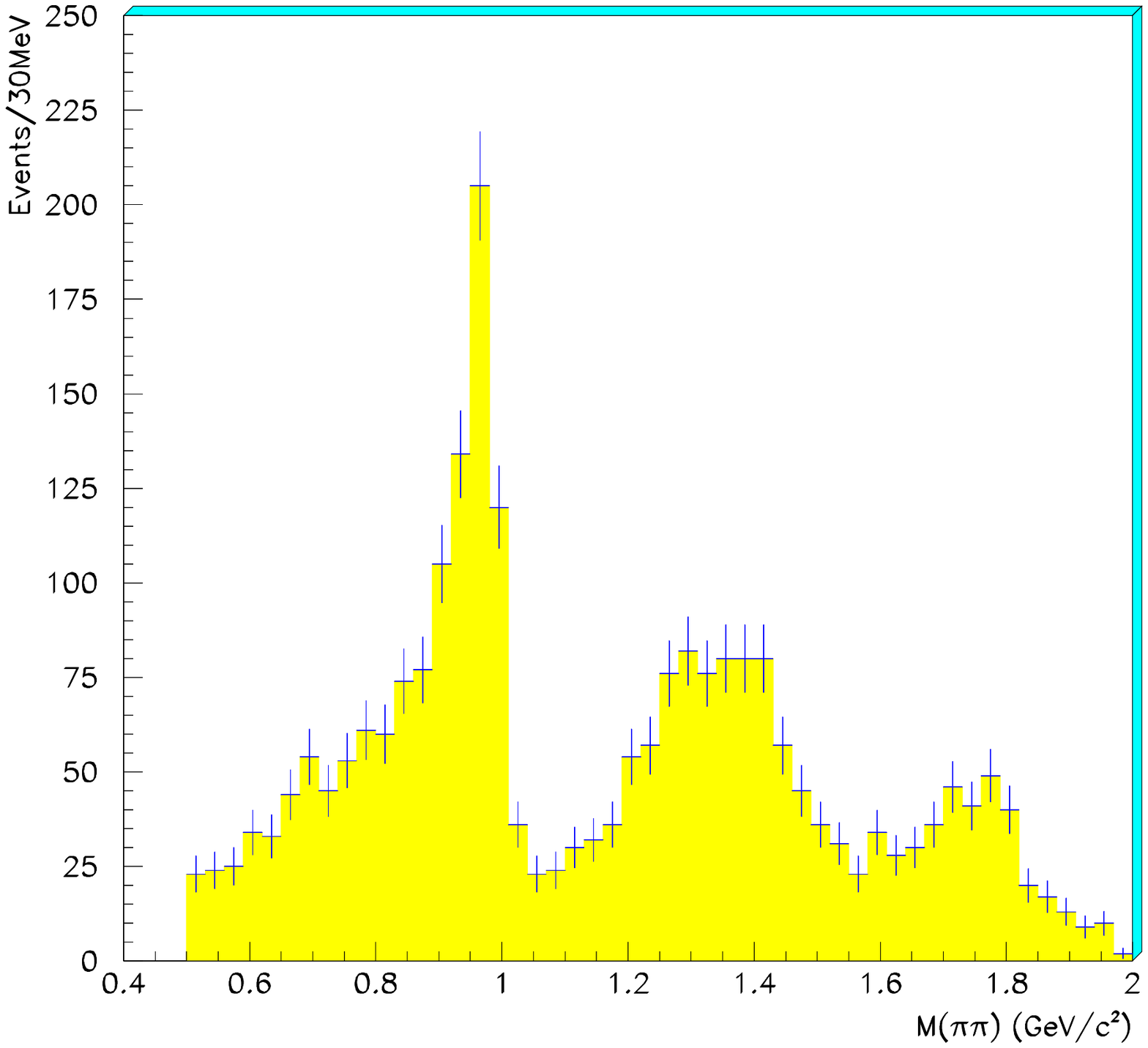,height=7cm,width=7cm}}
\vspace{-1.cm}\caption[]{The $\pi\pi$ mass of
$J/\Psi\to\phi\pi^+\pi^-$}
\label{fig-phi1}
\end{minipage}
\hspace{\fill}
\begin{minipage}[t]{75mm}
\centerline{\epsfig{file=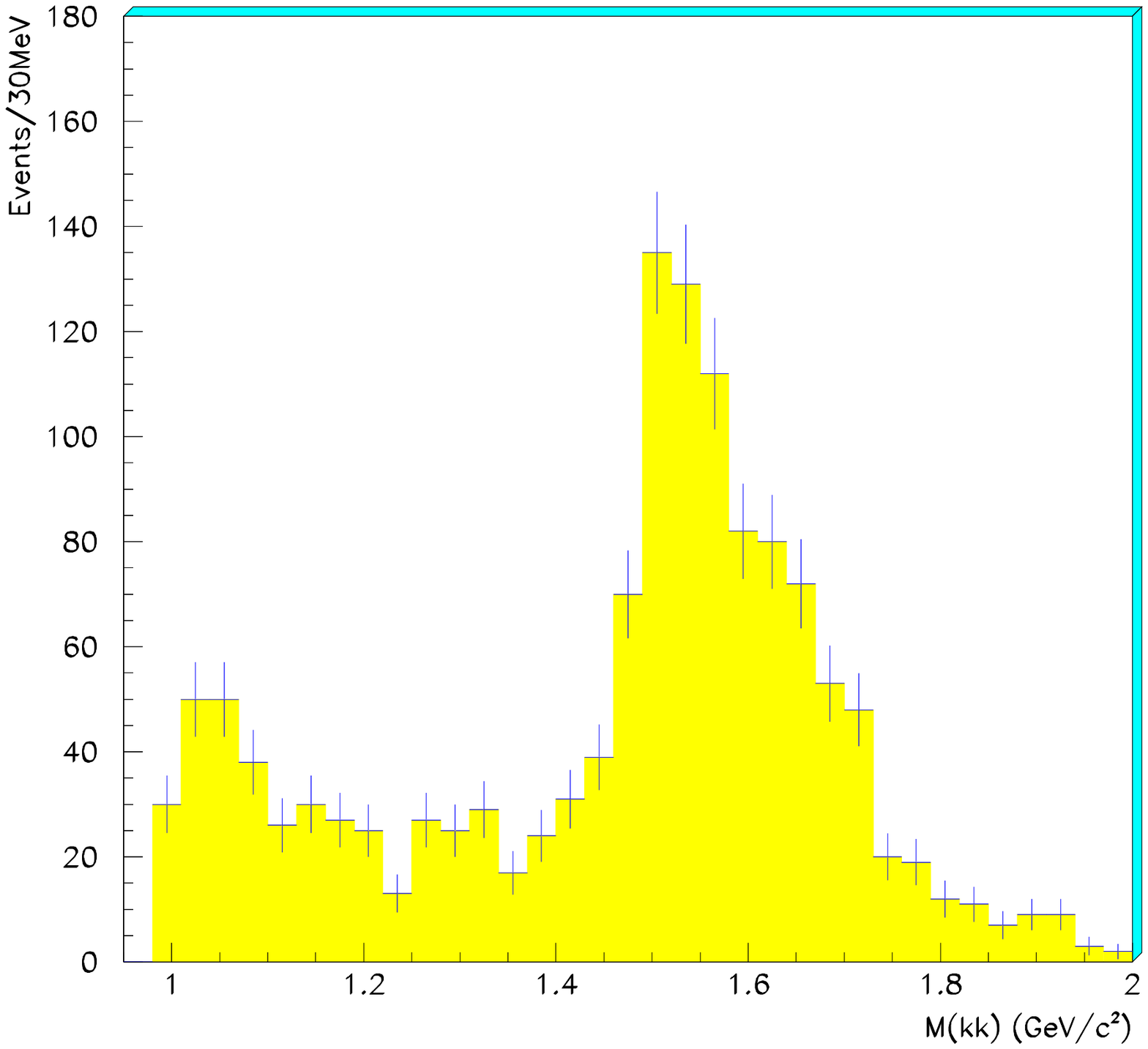,height=7cm,width=7cm}}
\vspace{-1.cm}\caption[]{The $K\bar K$ mass of $J/\Psi\to\phi K^+K^-$.}
\label{fig-phi2}
\end{minipage}  
\end{figure}

The $\pi\pi$ and $K\bar K$ invariant mass spectra for the $J/\Psi\to\phi
\pi^+\pi^-$ and $J/\Psi\to\phi K^+K^-$ based on the $2.3\times 10^7$
$J/\Psi$ events of BESII are shown in Fig.~\ref{fig-phi1} and
Fig.~\ref{fig-phi2}, respectively. Partial wave analyses are performed for
these two channels. Preliminary results\cite{Shen} indicate that 
(1) in the $\pi\pi$ mass spectrum of the $J/\Psi\to\phi \pi^+\pi^-$
process all three peaks at 980 MeV, 1330 MeV and 1770 MeV are dominantly 
$0^{++}$; (2) in the $K\bar K$ mass spectrum of the $J/\Psi\to\phi K^+K^-$ 
the peak at 1525 MeV is due to $f'_2(1525)$ while the $K\bar K$ threshold
enhancement and the shoulder around 1700 MeV are due to $f_0(980)$ and
$f_0(1710)$, respectively. The $f_0(1770)$ in the $J/\Psi\to\phi
\pi^+\pi^-$ and the $f_0(1710)$ in the $J/\Psi\to\phi K^+K^-$ seem to be
two separate resonances. The $f_0(1710\pm 20)\to K\bar K$ appears clearly
in the $J/\Psi\to\gamma K^+K^-$, $\omega K^+K^-$ and $\phi K^+K^-$;
while the $f_0(1760\pm 20)$ appears in
$J/\Psi\to\gamma\pi^+\pi^-\pi^+\pi^-$, $\omega\pi^+\pi^-\pi^+\pi^-$ and
$\phi\pi^+\pi^-$.

In summary, the $f_0(1710$-$1770)$ appears clearly in many
$J/\Psi\to\omega/\phi+X$ channels while $f_0(1500)$ is hardly visible in
any of these glueball-disfavored processes. The $\sigma$ and $f_2(1275)$
appear clearly only in the $J/\Psi\to\omega+X$ process, the $f'_2(1525)$
appears clearly only in the $J/\Psi\to\phi+X$ process, and $f_0(980)$ 
appears clearly in both processes.   

\section{$\Psi$ hadronic decays to baryons and anti-baryons} 

Baryons are the basic building blocks of our world. To understand the
internal quark-gluon structure of baryons is one of the most important
tasks in nowadays particle and nuclear physics.
From theoretical point of view,  since baryons represent the simplest
system in which the three colors of QCD neutralize into colorless objects
and the essential non-Abelian character of QCD is manifest, the systematic
study of various baryon spectroscopy will provide us with critical
insights into the nature of QCD in the confinement
domain\cite{Isgur,Klempt}.

$J/\Psi$ and $\Psi'$ decays provide an
excellent place for studying excited nucleons and hyperons -- $N^*$,
$\Lambda^*$, $\Sigma^*$ and $\Xi^*$ resonances. The corresponding Feynman
graph for the production of these excited
nucleons and hyperons is shown in Fig.~\ref{fig:3} where $\Psi$
represents either $J/\Psi$ or $\Psi'$.

\begin{figure}[htbp]
\vspace{-1.cm}
\hspace{1.0cm}\includegraphics[width=14cm,height=6cm]{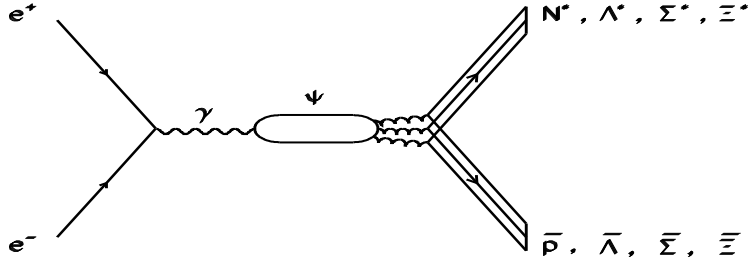}   
\vspace{-1.6cm}
\caption{$\bar pN^*$, $\bar\Lambda\Lambda^*$,
$\bar\Sigma\Sigma^*$ and $\bar\Xi\Xi^*$ production
from $e^+e^-$ collision through $\Psi$ meson.}
\label{fig:3}
\end{figure}

Since the vector charmonium $\Psi$ decays through three gluons and gluons
are flavor blind, the strange s quarks are produced at the same level as  
the non-strange u and d quarks.  Table~\ref{table:1} lists some interested
$J/\Psi$ decay branching ratios\cite{PDG}. The $p\bar p$,
$\Lambda\bar\Lambda$, $\Sigma^0\bar\Sigma^0$ and $\Xi\bar\Xi$ are indeed
produced at similar branching ratios. The branching ratios for
$\bar pN^*$, $\bar\Lambda\Lambda^*$, $\bar\Sigma\Sigma^*$ and
$\bar\Xi\Xi^*$ are expected to be of the same order of magnitude if one
ignores the phase space effect.
The $\bar\Omega\Omega^*$ channels have thresholds above or very close to
the mass of $\Psi'$ and cannot be studied here.

\begin{table}[htb]
\caption{ $J/\Psi$ decay branching ratios (BR$\times 10^3$) for some
interested channels\cite{PDG}}
\label{table:1}
\renewcommand{\arraystretch}{1.2} %enlarge line spacing
\begin{tabular}{ccccccc} 
\hline
$p\bar p$ & $\Lambda\bar\Lambda$ & $\Sigma^0\bar\Sigma^0$ & $\Xi\bar\Xi$
& $\Lambda\bar\Sigma^-\pi^+$ & $pK^-\bar\Lambda$ & $pK^-\bar\Sigma^0$\\
\hline
$2.1\pm 0.1$ & $1.4\pm 0.1$ & $1.3\pm 0.2$ & $1.8\pm 0.4$ &
$1.1\pm 0.1$ & $0.9\pm 0.2$ & $0.3\pm 0.1$\\
\hline
$p\bar n\pi^-$  & $p\bar p\pi^0$  &  $p\bar p\pi^+\pi^-$ &
$p\bar p\eta$ & $p\bar p\eta'$ & $p\bar p\omega$ &
$K^-\Lambda\bar\Xi^+$ ?\\
\hline
$2.0\pm 0.1$ & $1.1\pm 0.1$ & $6.0\pm 0.5$ & $2.1\pm 0.2$ &
$0.9\pm 0.4$ & $1.3\pm 0.3$ & $K^+\bar\Lambda\Xi^-$ ? \\
\hline
\end{tabular}\\
\end{table}

All channels listed in Table~\ref{table:1} are relative easy to be
reconstructed by BES. For example, for $K^-\Lambda\bar\Xi^+$, we can
select events containing $K^-$ and $\Lambda$ with $\Lambda\to p\pi^-$,
then from missing mass spectrum of $K^-\Lambda$ we should easily identify
the very narrow $\bar\Xi^+$ peak. The $K^-\Lambda\bar\Xi^+$ channel is a
very good place for studying $\Xi^*\to K\Lambda$. At present, not much is
known about $\Xi^*$ resonances\cite{PDG}. Only the ground $\Xi(1318)$
state and the first excitation state $\Xi^*(1530)$ are well established. 
There has not been a single new piece of data on $\Xi^*$ resonances since
PDG's 1988 edition. Various theoretical predictions by rather different
physical pictures\cite{Capstick,Glozman} are not challenged due to the
lack of data. With $J/\Psi$ and $\Psi'$ experiments at BEPC and upgraded
BEPCII in near future, we expect to complete the $\Xi^*$ resonance
spectrum as well as the $N^*$, $\Lambda^*$ and $\Sigma^*$ resonance
spectra. 

Among three-body channels listed in Table~\ref{table:1},
$\Lambda\bar\Sigma^-\pi^+$ and $pK^-\bar\Lambda$ can be used to study
$\Lambda^*\to\Sigma\pi$ and $NK$; $\Lambda\bar\Sigma^-\pi^+$
and $pK^-\bar\Sigma^0$ can be used to study
$\Sigma^*\to\Lambda\pi$ and $NK$; Channels containing p can be
used to study $N^*\to K\Lambda$, $K\Sigma$, $N\pi$, $N\pi\pi$, $N\eta$,
$N\eta'$ and $N\omega$. Many other channels not listed in
Table~\ref{table:1} can also be used to study these baryon
resonances.

In fact, the Feynman graph in Fig.~\ref{fig:3} is almost identical to
those describing the $N^*$ electro-production process if the direction of
the time axis is rotated by $90^o$. The only difference is that the
virtual photon here is time-like instead of space-like and couples to
$NN^*$ through a real vector charmonium meson $\Psi$. So all $N^*$
decay channels which are presently under investigation at
CEBAF(JLab, USA)\cite{Burkert}, ELSA(Bonn,Germany)\cite{Klempt}, 
GRAAL(Grenoble, France) and Spring8(KEK, Janpan) with   
real photon or space-like virtual photon can also
be studied at BEPC complementally with the time-like virtual photon.
In addition, for $\Psi\to\bar NN\pi$ and $\bar NN\pi\pi$, the $\pi N$
and $\pi\pi N$ systems are limited to be pure isospin $1/2$ due to isospin
conservation. This is a big advantage in studying $N^*$ resonances from
$\Psi$ decays, compared with $\pi N$ and $\gamma N$ experiments which
suffer difficulty on the isospin decomposition of $1/2$ and 
$3/2$\cite{Workman}. 

Based on 7.8 million $J/\Psi$ events collected at BEPC before 1996,
the events for $J/\Psi\to\bar pp\pi^0$ and $\bar pp\eta$ have been
selected and reconstructed with $\pi^0$ and $\eta$ detected in their
$\gamma\gamma$ decay mode\cite{Lihb}. For selected $J/\Psi\to\bar
pp\gamma\gamma$ events, the invariant mass spectrum of the $2\gamma$ is
shown in Fig.\ref{fig:4}. The $\pi^0$ and $\eta$ signals are clearly
there. The $p\pi^0$ invariant mass spectrum for $J/\Psi\to\bar
pp\pi^0$ is shown in Fig.~\ref{fig:5} with clear peaks around 1500 and 
1670 MeV. The $p\eta$ invariant mass spectrum for
$J/\Psi\to\bar pp\eta$ is shown in Fig.~\ref{fig:6} with clear enhancement
around the $p\eta$ threshold, peaks at 1540 and 1650 MeV; both have 
been determined to have $J^{P}={1\over 2}^-$ by a PWA analysis\cite{Lihb}.
From the relative branching ratio of $N^*(1535)$ to $\eta N$ and $\pi
N$\cite{PDG}, the narrow peak at 1500 MeV in $p\pi^0$ mass spectrum of 
$J/\Psi\to\bar pp\pi^0$ is expected to be mainly due to $N^*(1535)$.
With 23 million new $J/\Psi$ data collected by BESII in last few months,
many more channels, such as $p\bar n\pi^-$, $\bar pn\pi^+$, $\bar p\Lambda
K$, $\bar\Lambda K$, $\bar pp\omega$, $\bar pp\eta'$, $\bar pp\pi^+\pi^-$,
and $\bar pp\phi$ etc., are now under investigation. 

\begin{figure}[htb] % fig.4,5,6
\begin{minipage}[t]{50mm}
\centerline{\epsfig{file=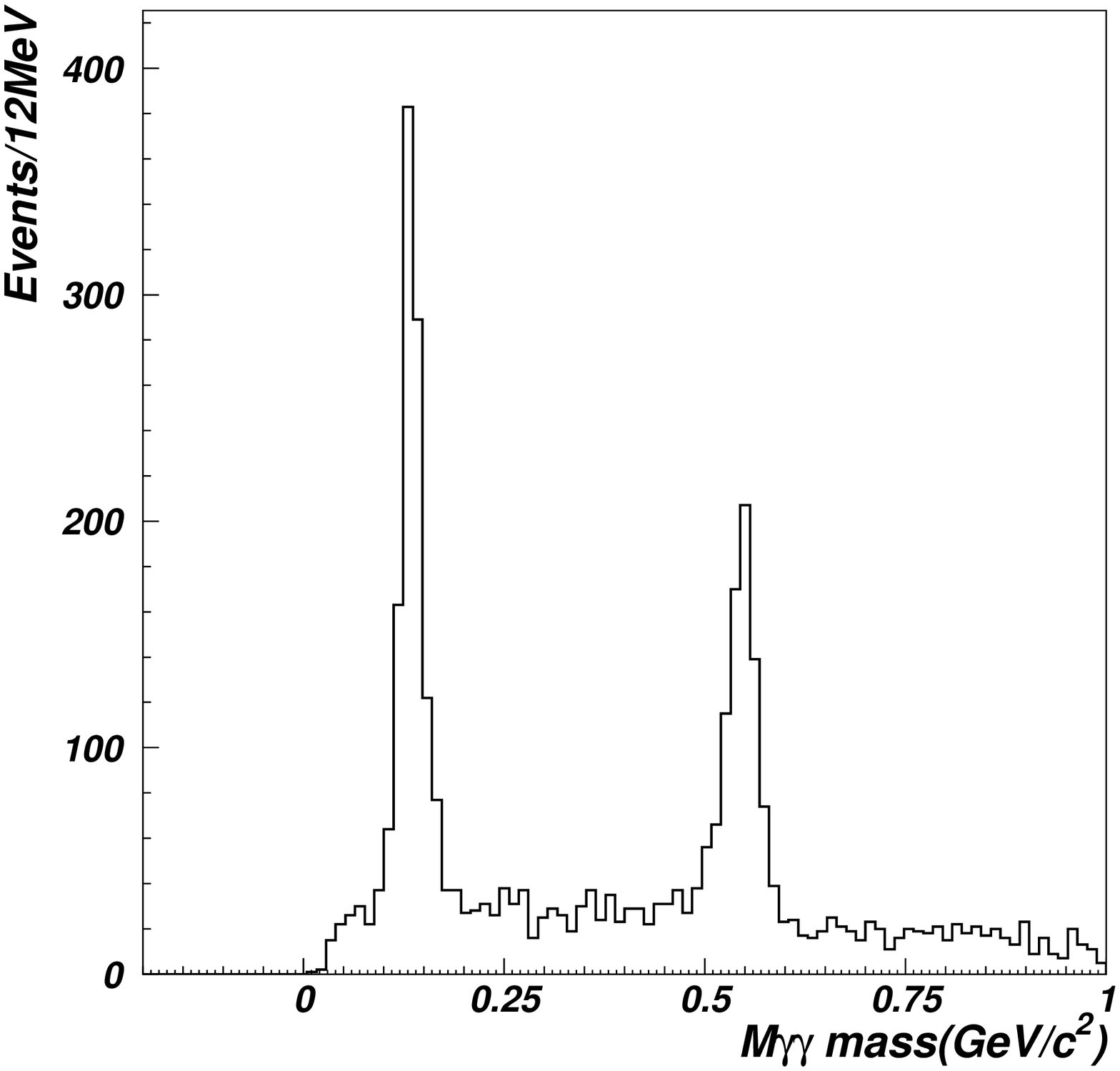,height=2.in,width=2.in}}
\vspace{-1.cm}\caption[]{$\gamma \gamma$ invariant mass spectrum after 4C
fit for $J/\Psi\to\bar pp\gamma\gamma$}
\label{fig:4}
\end{minipage}
\hspace{\fill}
\begin{minipage}[t]{50mm}
\centerline{\epsfig{file=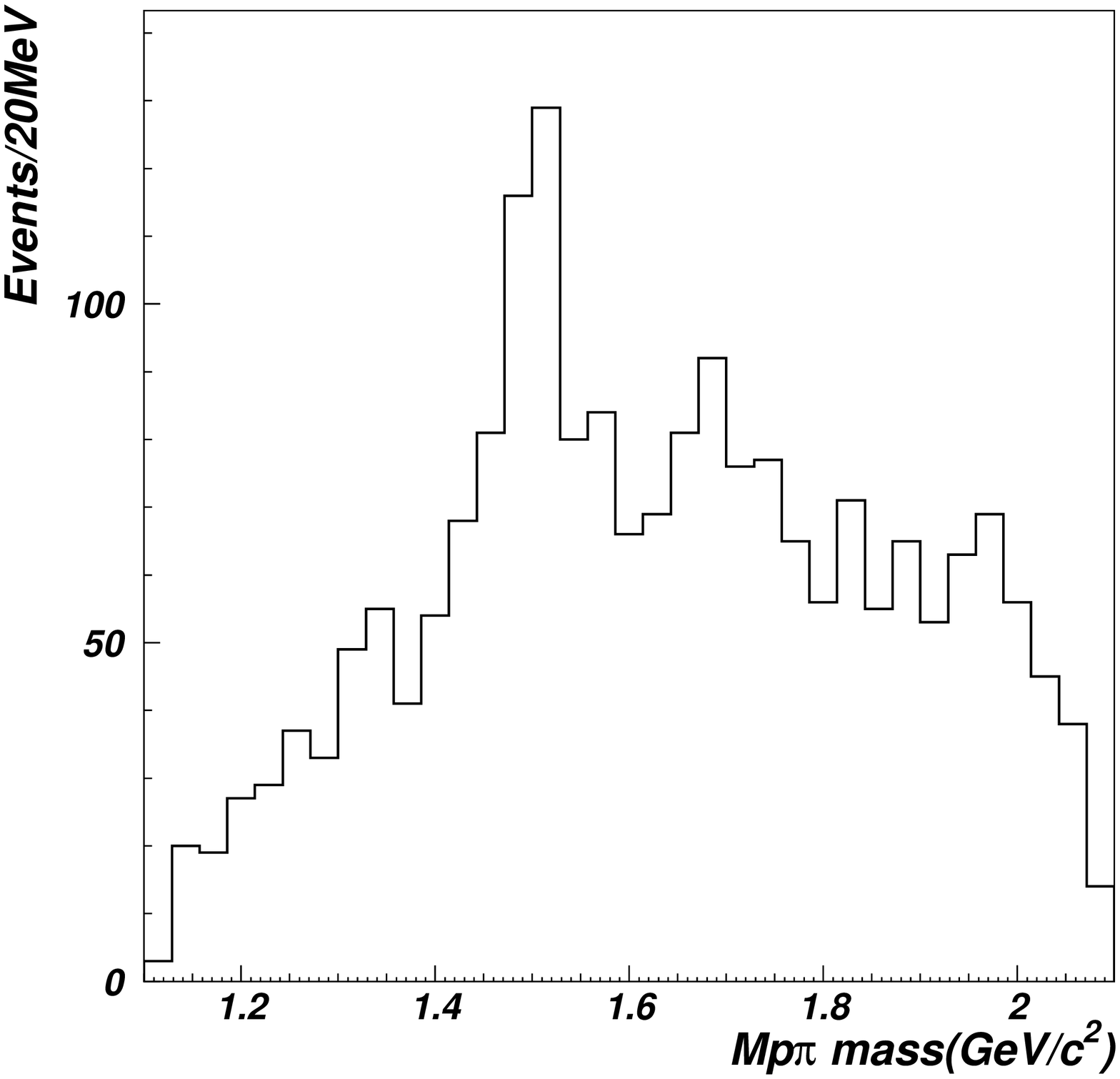,height=2.in,width=2.in}}
\vspace{-1.cm}
\caption[]{$p\pi^0$ invariant mass spectrum for
$J/\Psi\to\bar pp\pi^0$.}
\label{fig:5}
\end{minipage}  
\hspace{\fill}
\begin{minipage}[t]{45mm}
\centerline{\epsfig{file=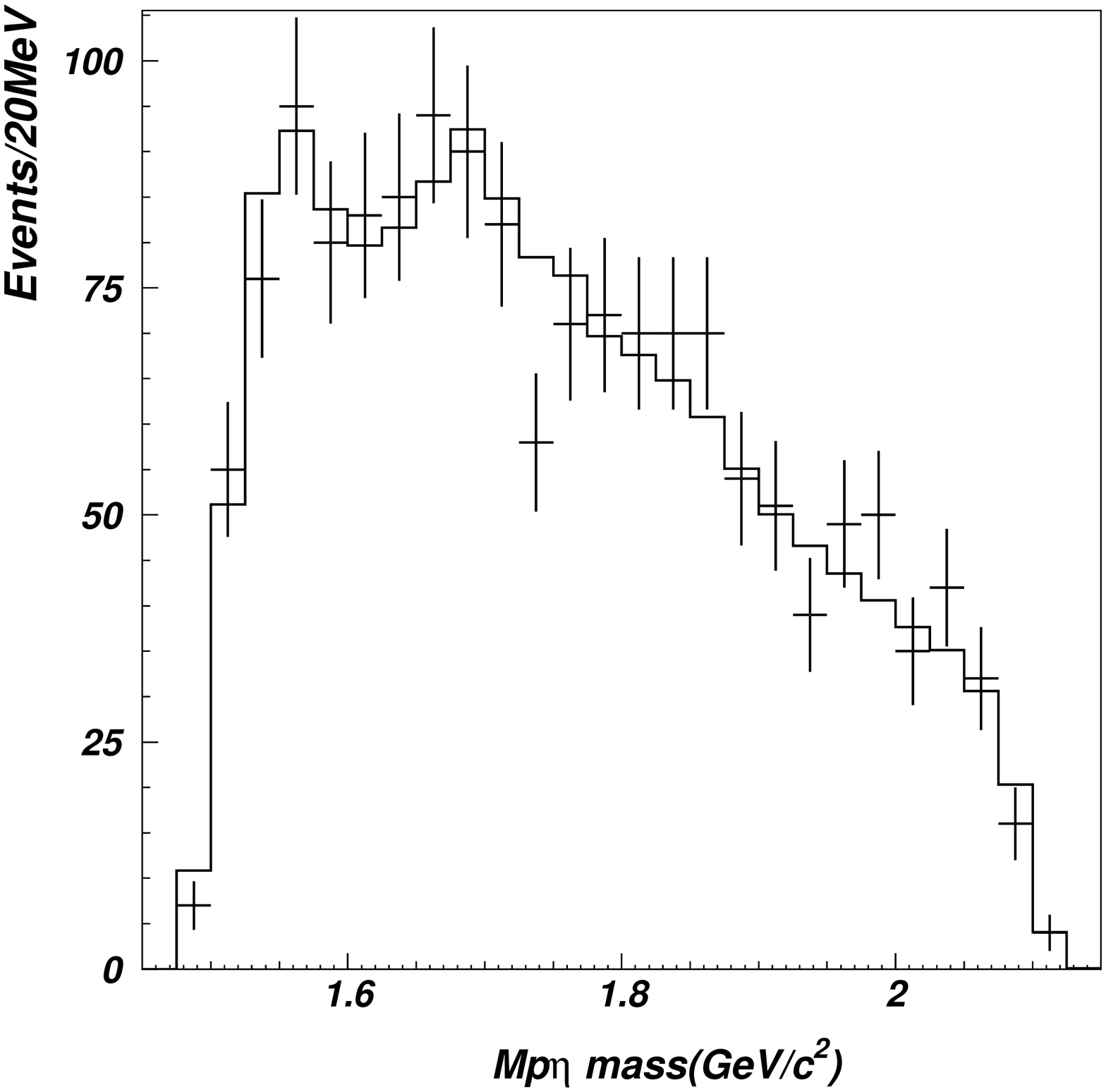,height=2.in,width=2.in}}
\vspace{-1.cm}
\caption{$p\eta$ invariant mass spectrum for $J/\Psi\to\bar
pp\eta$.}
\label{fig:6}
\end{minipage}  
\end{figure}

On theoretical side, the coupling of $\Psi\to\bar pN^*$ 
provides a new way to probe the internal quark-gluon structure of the
$N^*$ resonances\cite{Zou1}. In the simple
three-quark picture of baryons, as shown in Fig.~\ref{fig:3}, three
quark-antiquark pairs are created independently via a symmetric
three-gluon intermediate state with no extra interaction other than the
recombination process in the final state to form baryons. This is quite
different from the mechanism underlying the $N^*$ production from the
$\gamma p$ process where the photon couples to only one quark and
unsymmetric configuration of quarks is favored.  Therefore the processes
$\Psi\to\bar pN^*$ and $\gamma p\to N^*$ should probe different aspects of
the quark distributions inside baryons.
Since the $\Psi$ decay is a glue-rich process, it is also regarded as a
good place for looking for hybrid $N^*$\cite{Page}.

In summary, the $J/\Psi$ and $\Psi'$ experiments at BEPC provide an
excellent place for studying excited nucleons and hyperons -- $N^*$,
$\Lambda^*$, $\Sigma^*$ and $\Xi^*$ resonances. 
The completion of the light quark (u,d,d) baryon spectroscopy is of
crucial importance for us to reveal quark gluon structure of matter.

\section{Outlook}

All PWA results summarized in this paper are based on 7.8 million $J/\Psi$
events collected by the old version of the BES detector (BESI) before
1996. Since November 1999, 23 million new $J/\Psi$ events have been
collected with the new improved BES detector (BESII) and 27 million more
$J/\Psi$ events are going to be taken before next May.

A major upgrading of the collider to the BEPC2 has been approved
by the Chinese central government very recently. A further one order of
magnitude more statistics is expected to be achieved. Such statistics will
enable us to perform partial wave analyses of plenty important channels
for both meson spectroscopy and baryon spectroscopy from the $J/\Psi$ and
$\Psi'$ decays.
We expect BEPC2 to play a very important unique role in many aspects
of light hadron spectroscopy, such as hunting for the glueballs and
hybrids, extracting $u\bar u+d\bar d$ and $s\bar s$ components of mesons,
and studying excited nucleons and hyperons, {i.e.}, $N^*$, $\Lambda^*$,
$\Sigma^*$ and $\Xi^*$ resonances.

\bigskip
{\bf Acknowledgements:} It is a pleasure to thank Prof. Martin Faessler
and his staff for the successful organization of LEAP2000 on the nice
Venice island and the invitation. Deep thanks to my collaborators in the
BES experiment at BEPC.

\end{document}